\documentclass[pre,reprint,superscriptaddress]{revtex4-1}

\usepackage{amsmath}
\usepackage{amssymb}
\usepackage{graphicx}

\usepackage{textcomp}

\begin{document}
\title{Comparative statistical study of two local clustering coefficient formulations as tropical cyclone markers for climate networks}

\author{Mikhail Krivonosov}
\affiliation{Lobachevsky State University of Nizhny Novgorod, Nizhny Novgorod, 
	Russia}
\author{Olga Vershinina}
\affiliation{Lobachevsky State University of Nizhny Novgorod, Nizhny Novgorod, 
	Russia}
\author{Anna Pirova}
\affiliation{Lobachevsky State University of Nizhny Novgorod, Nizhny Novgorod, 
	Russia}
\author{Shraddha Gupta}
\affiliation{Potsdam Institute for Climate Impact Research, PO Box 601203, 14412 Potsdam, Germany}
\affiliation{Department of Physics, Humboldt University, Berlin, Germany}
\author{Oleg Kanakov}
\affiliation{Lobachevsky State University of Nizhny Novgorod, Nizhny Novgorod, 
	Russia}
\author{J\"urgen Kurths}
\affiliation{Lobachevsky State University of Nizhny Novgorod, Nizhny Novgorod, Russia}
\affiliation{Potsdam Institute for Climate Impact Research, PO Box 601203, 14412 Potsdam, Germany}
\affiliation{Department of Physics, Humboldt University, Berlin, Germany}
		
\begin{abstract}
We introduce a new formulation of local clustering coefficient for weighted correlation networks. This new formulation is based upon a definition introduced previously in the neuroscience context and aimed at compensating for spurious correlations caused by indirect interactions. We modify this definition further by replacing Pearson's pairwise correlation coefficients and three-way partial correlation coefficients by the respective Kendall's rank correlations. This reduces statistical sample size requirements to compute the correlations, which translates into the possibility of using shorter time windows and hence into a shorter response time of the real-time climate network analysis. We construct evolving climate networks of mean sea level pressure fluctuations and analyze anomalies of local clustering coefficient in these networks. We develop a broadly applicable statistical methodology to study association between spatially inhomogeneous georeferenced multivariate time series and binary-valued spatiotemporal data (or other data reducible to this representation) and use it to compare the newly proposed formulation of local clustering coefficient (for weighted correlation networks) to the conventional one (for unweighted graphs) in terms of the association of these measures in climate networks to tropical cyclones. Thus we substantiate the previously made observation that tropical cyclones are associated with anomalously high values of local clustering coefficient, and confirm that the new formulation shows a stronger association.
\end{abstract}

\maketitle
	

\section{Introduction}
Network analysis is recognized nowadays as a powerful tool in complex system studies, including climate science. In essence, it is a specialization of the general correlation networks approach \cite{langfelder2008wgcna}, which is also well established in systems biology \cite{bruggeman2007nature, van2007art, friedman2012inferring, rosato2018correlation} and neuroscience \cite{fransson2008precuneus, bassett2017network}. Correlation networks in climate science are constructed from cross-correlations in multivariate time series of chosen climate variables between nodes of a spatial grid (e.g. a geographic coordinate grid on Earth surface). Such networks are then used to compute various graph metrics, which in turn become subject to subsequent analysis aimed at identifying patterns of climate dynamics \cite{tsonis2006networks, donges2009complex, gozolchiani2011emergence}.

Tropical cyclones (TCs) are extreme weather events that can cause tremendous destruction to life and property. Accurate forecasting and analysis of TCs and their tracks remains a challenge and a major area of research in climate science, even today. Climate networks, which have been widely used to study different weather dynamics, were recently employed in the study of extreme rainfall events associated with TCs \cite{ozturk2018complex, ozturk2019network, traxl2016size}. In \cite{gupta2021complex}, the authors have undertaken a time-evolving functional network approach to capture the rapid spatio-temporal evolution of the network topology of the regional weather system associated with a TC. They propose suitable network metrics to detect TC tracks from mean sea level pressure (MSLP) data and study their dynamics.

It is essential for the efficiency of this approach that as much as possible of information contained in the initial data are retained in the feature set produced by network analysis. Climate networks are typically (including those studied in \cite{gupta2021complex}) unweighted graphs obtained by thresholding the correlation matrix of the input data. Network nodes are associated to the variables in the input data; in case of a single climatic observable, e.g. MSLP, and, hence, in a single-layer network, the network nodes are in one-to-one correspondence with the spatial grid nodes on the Earth surface. A link between two specific nodes in the network is assumed to exist if the corresponding component of the correlation matrix (i.e. the  correlation coefficient in this specific pair of variables) exceeds a chosen threshold. The threshold value is a free parameter of the method, and generally allows tuning to maximize performance (in whatever quantitative sense), but it is also common to fix the threshold value by specifying the edge density (defined as the fraction which linked pairs of nodes constitute among the total number of node pairs), e.g. at 5\% or 10\%, without any further optimization. Regardless of whether the threshold value is optimized or not, thresholding inevitably leads to information loss.

A known way to mitigate the information loss introduced by thresholding is to construct and analyze an ensemble of several unweighted graphs at once, by using a series of threshold values; this approach has been successfully applied to genetic network analysis \cite{bockmayr2013new}. It is also possible to eliminate the thresholding operation at all, by analyzing a full weighted graph whose edge weights are determined by (in the simplest case, taken equal to) the respective pairwise correlation coefficients. The transition from unweighted to weighted networks calls for a proper extension of graph metrics definitions. For many metrics such extensions are available \cite{Newman2016}, but may be not unique, thus giving rise to an additional problem of choosing the best weighted-network modification of a particular graph metric. An important consideration to guide this choice is that the networks of interest are, in essence, correlation networks.

The weighted-network approach has been successfully implemented in the neuroscience context \cite{frontiers2018}, where an improved formulation of local clustering coefficient (LCC) specially focused on weighted correlation networks was proposed and demonstrated to outperform the conventional thresholding approach in terms of an illustrative neuroscience problem (revealing the age dependence of human brain network structure based on functional magnetic resonance data). The key distinctive feature of the special definition of LCC for weighted correlation networks \cite{frontiers2018} consists in accounting for three-way partial correlations in order to compensate for spurious correlations caused by indirect interactions, which otherwise disguise the true interaction structure in the correlation matrix.

No literature is currently available to implement this approach in climate science. The present study aims at filling this gap by adapting the correlations-focused weighted-network LCC definition from \cite{frontiers2018} to climate network analysis. We further modify the LCC definition of \cite{frontiers2018} by switching from (parametric) Pearson's to (non-parametric) Kendall's correlations (which moves from capturing linear to more general monotonic dependencies between variables and imposes weaker requirements to sample size as compared to both Pearson's and Spearman's correlation \cite{bonett2000sample}) and compute this new LCC formulation for MSLP correlation networks. We find and quantify the relation of thus computed LCC to available data on the trajectories of tropical cyclones, thereby supporting and extending the results of \cite{gupta2021complex}, where LCC anomalies of MSLP correlation networks were discovered in the vicinity of TCs. We show that the LCC modification based on three-way partial correlations outperforms the unweighted thresholded network LCC as a marker of tropical cyclone. We expect this new measure to be especially effective for the analysis of strongly time-dependent climate phenomena.

\section{Methods}
\subsection{Terminology, conventions and notation}\label{sec_termin}
\textit{Attribute} will mean here any physical quantity which is measurable directly or indirectly at a specific time and location (e.g. MSLP, see Sec.~\ref{sec_srcdata}), or a computable quantity which is also attributed to a specific time and location (e.g. LCC of an evolving correlation network, see Sec.~\ref{sec_LCC}).

\textit{Observation} is an individual value of an attribute (obtained by measurement or computation).

\textit{Time grid} is a set of moments in time at which observations are made. Here the time grid $\{t_m\}$ is equidistant (periodic), and its period is referred to as \textit{sampling period}. 

\textit{Spatial grid} is a set of fixed locations in space at which observations are made. We use a spatial grid of ERA5 single-level climate data \cite{hersbach2020era5} (see details in Sec.~\ref{sec_srcdata}), which has no altitude dimension (MSLP data are attributed to the mean sea level by definition), and the grid is periodic in geographic coordinates (lattitude and longitude), but not equidistant on the Earth surface. Since we use a single attribute (specifically, MSLP) to construct correlation networks (see Sec.~\ref{sec_evonet}), the nodes of these networks are by construction in one-to-one correspondence to the spatial grid nodes; such networks are referred to as single-layer networks. The quantity of spatial grid nodes is denoted as $N$.

\textit{Space-time grid} is the Cartesian product of the spatial grid and the time grid.

\textit{Time series} is a sequence of observations made at consecutive points of the time grid. \textit{Univariate} time series contain a single observation per each moment of time, and \textit{multivariate} consist here of simultaneous observations of a single attribute (e.g. MSLP or LCC) at multiple locations (space grid nodes). Generally, multivariate time series may as well encompass several attributes, giving rise to multi-layer correlation networks.

Subscript indices $i$, $j$, $l$ will label the spatial grid nodes (and, equivalently, the correlation network nodes); superscript bracketed indices $(m)$, $(n)$ will label the time grid points. For example, the observation of an attribute $v$ at the spatial node $i$ at the time $t_m$ will be denoted as $v_i(t_m)$, or, equivalently, as $v_i^{(m)}$; notations with braces (e.g. $\{v_i^{(m)}\}$) mean the entire time series.

\textit{Variate} generally means a random quantity, but here it will also refer to a component of multivariate time series, when described in terms of probability theory. A notation for such a variate (e.g. $v_i$) is obtained by omitting the temporal index from the notation of the observations in the series. 

\textit{Sample} of a variate is generally a set of its observations (not necessarily time-stamped or even ordered); here, a sample is always represented by a time series. It is common in statistics to assume that the observations in a sample are independent and identically distributed, but this is not the case within our problem statement. Here we treat all time series under study as realizations of discrete-time stochastic processes \cite[Chap.~9]{papoulis2002probability}. Stationarity and ergodicity of these underlying processes are important issues to consider in order to treat a time series as a statistical sample (see Sec.~\ref{sec_ecdf}, \ref{sec_assoc}).

\textit{Georeferenced time series} means multivariate time series, where each component (variate) is attributed to a specific location in space, which does not change over time; these locations are nodes of the spatial grid.

\textit{Trajectory} is another form of spatiotemporal data, where the location of a moving object (here, TC center) is recorded over a time grid; these locations are not assumed to coincide with spatial grid nodes.

Our study involves several statistical and graph-theoretic methods, which are applied in sequence. Further, we provide a general description for each method, and specify how the method is applied within our computation chain. \textit{General input data} in a method's description mean the type of data to which the method is generically applicable; \textit{actual input data} mean the data to which the method is actually applied in our computation.

We study association (dependence) between processed georeferenced time series (here, that of LCC in two different formulations) on the one hand, and TCs (represented by a set of available TC center trajectories) on the other. For generality, we refer to a particular attribute under this association study (here, to each particular formulation of LCC) as \textit{indicator}, and to a class of events (here, TCs) which occur in space and time and may be represented as time series of a binary valued (taking on values `yes' or `no', see Sec.~\ref{sec_assoc}) attribute over the space-time grid as \textit{event of interest}. Moreover, prior to a statistical analysis for association, the indicator time series are transformed (regularized) so as to provide that all variates within the transformed series (each variate pertaining to a specific node of the spatial grid) are characterized by the same universal probability distribution (namely, uniform distribution on $[0,1]$, see Sec.~\ref{sec_ecdf}). The outcome of this regularization will be referred to as \textit{transformed indicator}. The association analysis methodology described in Sec.~\ref{sec_ecdf}--\ref{sec_assoc} may be applied to any indicator and event of interest, provided they meet the assumptions outlined in the respective methods descriptions.

\subsection{Cimate data}\label{sec_srcdata}
As a source of initial data (in the form of georeferenced multivariate time series) for constructing climate correlation networks, we use the ERA5 reanalysis data \cite{hersbach2020era5}, which is essentially a model-based interpolation of available climatic observational data to a regular grid over geographic coordinates and time. The attribute under study is mean sea level pressure (MSLP), i.e. the calculated value of local atmospheric pressure at the standard (long-term average) local sea level. We use time series of MSLP on a coordinate grid with step of 0.75\textdegree{} over latitude and longitude, and time grid with sampling period of 3 hours. The data are taken for the sea surface (land excluded) within the range 4.5\textdegree N to 30.75\textdegree N latitude and 49.5\textdegree E to 100.5\textdegree E longitude (northern part of the Indian Ocean), over the time range 00:00:00 01.01.1982 to 21:00:00 31.12.2020, thus the time grid consists of $M_{\text{source}}$=113960 points.

As long as correlation networks are meant to characterize fluctuations of the variates under study, it is beneficial to suppress the deterministic component in the time series prior to the correlation analysis. For that sake, following the approach of \cite{gupta2021complex}, we preprocess the original ERA5 data by subtracting the respective local daily climate normals for each location (spatial grid node) and each date of year, which are computed by averaging the attribute (here, MSLP) at the specific spatial grid node over all observations for the particular date of year and over all years of observation (for simplicity, data for 29 February of leap years are discarded). This eliminates the annual cycle from each component of the time series. The result of this subtraction, further referred to as \textit{MSLP anomaly} and denoted $\{v_i^{(m)}\}$, constitutes georeferenced time series on the same space and time grids as the original ERA5 data. The MSLP anomaly data are subject to the subsequent correlation analysis (see Sec.~\ref{sec:kendall}, \ref{sec_evonet}).

The other type of spatiotemporal data in our study are the Best Track data \cite{mohapatra2012best} on tropical cyclones in the region, which are produced and published by the Regional Specialized Meteorological Centre for Tropical Cyclones Over North Indian Ocean (India) and contain information on the position and strength of all registered tropical cyclones in the region over time. Currently, we ignore the TC strength information and use only TC center trajectories (see Sec.~\ref{sec_assoc}).

\subsection{Kendall's correlation}\label{sec:kendall}
To quantify correlation, in line with \cite{gupta2021complex}, we use the Kendall's rank correlation coefficient $\tau$ \cite{kendall1938new} (more precisely, the Goodman-Kruskal modification thereof, see below). \textit{General input data} for computing a correlation coefficient of two random variates is their joint bivariate sample. We compute pairwise correlation coefficients $\tau_{ij}$ for the respective pairs of MSLP anomalies $v_i$ and $v_j$ (pertaining to the spatial grid nodes $i$ and $j$). The \textit{actual input data} for obtaining a coefficient $\tau_{ij}$ is a joint sample of the variates $v_i$ and $v_j$ constructed from a section of the multivariate time series of MSLP anomalies taken within a sliding time window (see Sec.~\ref{sec_evonet}).

Given a bivariate sample consisting of $w$ joint observations $\{(v_i^{(k)}, v_j^{(k)})\}_{k=1..w}$, the Kendall's rank correlation coefficient $\tau_{ij}$ is defined as a normalized difference between the counts of concordant and discordant pairs, i.e. pairwise combinations of bivariate observations from the sample. Namely, a pair of bivariate observations $\big((v_i^{(m)}$, $v_j^{(m)})$, $(v_i^{(n)}$, $v_j^{(n)})\big)$ is called concordant (discordant), if the product $(v_i^{(m)}-v_i^{(n)}) \cdot (v_j^{(m)}-v_j^{(n)})$ is positive (negative), implying that these particular observations show an increasing (decreasing) dependence between $v_i$ and $v_j$. Normalization ensures that $\tau_{ij}\in[-1,1]$; in particular, $\tau_{ij}=1$ ($\tau_{ij}=-1$) implies that all observations within the sample are compliant with some deterministic increasing (decreasing) functional dependence between $v_i$ and $v_j$. Different formulations for the normalizing denominator are available in the literature, depending on the chosen way to account for ties (i.e. equality cases with $v_i^{(m)}=v_i^{(n)}$ or $v_j^{(m)}=v_j^{(n)}$), which is still a matter of research \cite{amerise2015correction}. As long as ties in our problem have a negligible impact due to the continuous nature of the climate variables under study, the method of their resolution is not actually important. For definiteness, we discard tied pairs if such occur, and define the normalizing denominator as the total of concordant and discordant pairs (ties excluded), as suggested in \cite{adler1957modification}, \cite[Eq.~(15.2) and below therein]{kruskal1958ordinal}, the resultant quantity also known as the Goodman-Kruskal gamma coefficient \cite[Eq.~(2.39)]{gibbons2003nonparametric}, which is named after the authors who applied it to variates taking on finite sets of values \cite[Eq.~(21)]{goodman1954measures}.

Our choice of the Kendall's rank correlation coefficient over other available measures of association is due to the following considerations: (i) it imposes weaker requirements on the sample size $w$ as compared to both Pearson's and Spearman's correlation coefficients \cite{bonett2000sample}, which translates into a shorter length of the sliding time window (see Sec.~\ref{sec_evonet}), and hence a better time resolution; (ii) like any rank-order statistics (including Spearman's correlation), it is invariant to any monotone nonlinear transformations of variables \cite[Sec.~5.5]{gibbons2003nonparametric}, which is not the case with Pearson's correlation being only invariant to linear transformations;
(iii) it is more robust to outliers (large-amplitude noise) than both Pearson's and Spearman's correlation coefficients \cite{xu2013comparative}.

\subsection{Evolving correlation networks}\label{sec_evonet}
Here we describe a method of constructing a time sequence of correlation networks, which is referred to as an \textit{evolving network}. This is a modification of the approach taken in \cite{gupta2021complex}. \textit{General input data} can be any multivariate time series; the \textit{actual input data} are the georeferenced time series of MSLP anomaly $\{v_i^{(m)}\}$ (see Sec.~\ref{sec_srcdata}).

The full symmetric matrix of Kendall's pairwise correlation coefficients $(\tau_{ij})$ is computed for input multivariate time series over a sliding time window of fixed length $w$ points of the time grid, and attributed to the ending time of the window. We take $w=16$; given the sampling period of 3 hours, the window length amounts to 2 days. We performed computations with greater window sizes, but they produce weaker association of LCC anomalies to TCs, which is explained by a worse time resolution, i.e. slower response time of the correlation network to the processes associated with TCs.

Namely, for each point of the time grid of the input data, starting from the $w$th and onwards (further referred to as the ``current time''), we use a short section (``window'') of the input time series consisting of $w$ consecutive joint observations, of which the latest is at the current time, as a sample to compute all pairwise correlation coefficients $\tau_{ij}$ between the variates (as described in Sec.~\ref{sec:kendall}).

We consider both weighted and unweighted correlation networks, the former defined by the link weight matrix taken equal to the Kendall's correlation matrix $(\tau_{ij})$, and the latter constructed by thresholding the correlation matrix: a pair of nodes are considered connected when their correlation coefficient exceeds a threshold, which in turn is chosen so that the fraction of connected nodes among the total number of node pairs (edge density) is a specified quantity (taken equal to 5\% and 10\% in the computations). Thus, the adjacency matrix $(a_{ij})$ of an unweighted network (by definition, $a_{ij}=1$ if a link is present between the nodes $i$ and $j$, and $a_{ij}=0$ otherwise) is expressed as
\begin{equation}\label{eq_thr}
a_{ij}=1(\tau_{ij}>\theta),
\end{equation}
where $\theta$ is the threshold value; the notation $1(\tau_{ij}>\theta)$ is defined to be unity whenever the condition in the brackets holds, and zero otherwise. As soon as the correlation matrices are symmetric, all correlation networks under consideration are undirected graphs.

The procedure of constructing correlation networks is repeated with the current time advanced sequentially to each further time grid point, and the time window shifted accordingly \footnote{We developed an optimized algorithm to compute Kendall's rank correlation coefficient over a sliding time window, which benefits from reusing the computation results obtained at the previous positions of the window.}. This way, we obtain a sequence of networks, each attributed to the respective current time, which can be seen as instances of a time-evolving network. The time interval between consecutive instances of this evolving network equals the sampling period of the input time series. The chosen positioning of the current time within the time window (at the ending point of the window) ensures causality of our analysis, which means that computing any quantity attributed to the current time never requires the knowledge of any data which are in the future relative to this current time. The chosen time window length of $w=16$ observations (or 2 days) is a reasonable trade-off between being too short for the Kendall's coefficient to be meaningful, and being too long to respond quickly enough to the TC-associated changes in the climate system.

The resultant evolving networks are represented by a time sequence of weight matrices $\{(\tau_{ij})^{(m)}\}$ for the weighted network, and adjacency matrices $\{(a_{ij})^{(m)}\}$ for the unweighted network. Since the earliest instance of a correlation network is attributed to the $w$th time grid point of the source time series, the time grid of the evolving network lacks the $w-1$ initial points of the source time grid; the total number of network instances $M$ is accordingly smaller than the length of the source ERA5 time series $M_{\text{source}}$:
\begin{equation}\label{eq_Ntime}
	M = M_{\text{source}} - w + 1.
\end{equation}
Since the time series of network measures (see Sec.~\ref{sec_LCC}) obtained from an evolving network inherit its time grid, this grid (of length $M$, starting at the $w$th point of the source grid) is always implied below.


\subsection{Local clustering coefficient}\label{sec_LCC}
This Section describes the procedure of computing LCC in correlation networks, which can be unweighted or weighted. \textit{General input data} for defining and computing LCC in an unweighted (weighted) undirected graph can consist of any representation of the graph, e.g. the quantity of nodes $N$ and the adjacency matrix $(a_{ij})$ (link weight matrix $(\tau_{ij})$). The \textit{actual input data} are the instances of the evolving correlation networks $\{(a_{ij})^{(m)}\}$, $\{(\tau_{ij})^{(m)}\}$ obtained from the procedure described in Sec.~\ref{sec_evonet}. Further we omit the temporal index $(m)$, assuming that LCC computation is performed in turn for each instance of an evolving network. Since the graphs under study are essentially correlation networks, they admit the use of specially defined LCC for weighted correlation networks \cite{frontiers2018}, as described below.

\begin{figure}
	\centering
	$A$\includegraphics[width=0.4\columnwidth]{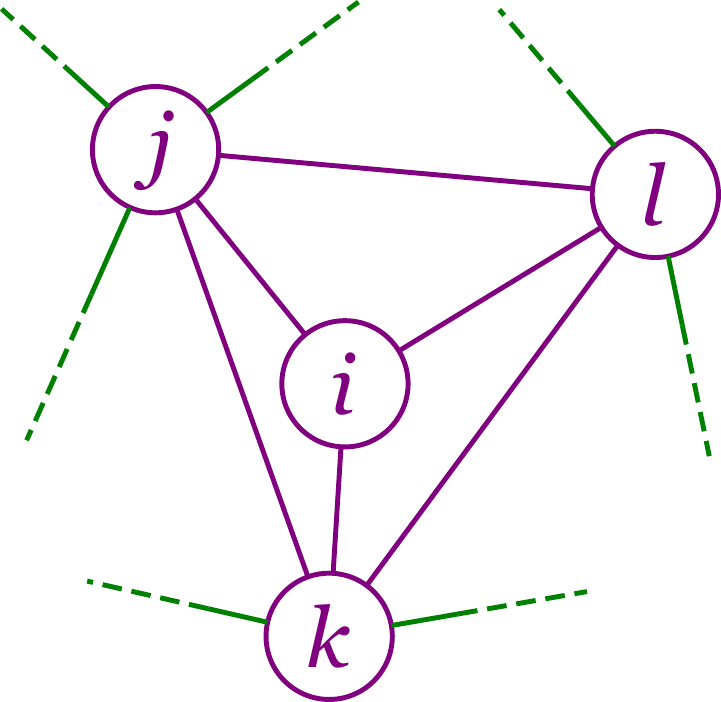}\hfill
	$B$\includegraphics[width=0.4\columnwidth]{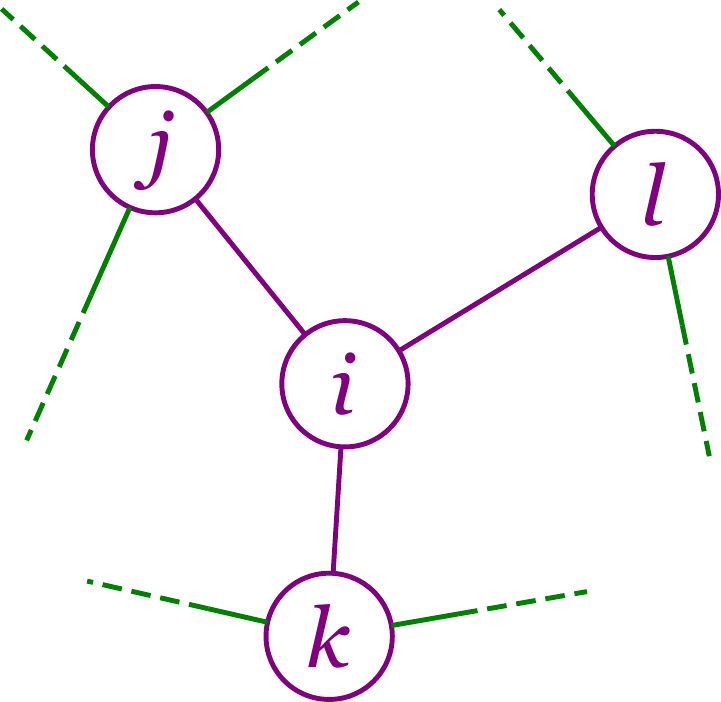}
	\caption{Sample network fragments illustrating LCC values of 1 and 0. Labeled circles and solid lines shown in magenta are the nodes and the edges of the subgraph which determines LCC of the node $i$. Green lines with dashed endings denote insignificant connections to the rest of the network. Panel $A$: node $i$ together with its direct neighbors constitute a clique, and thus LCC of this node equals 1 (all neighbors of $i$ are also neighbors to each other). Panel $B$: node $i$ with its neighbors constitute a star subgraph, and LCC of $i$ equals 0 (none of the neighbors of $i$ are neighbors of each other).}\label{fig_net_samples}
\end{figure}

The local clustering coefficient for a given node $i$ in a general unweighted undirected graph is defined as the ratio of the number of closed triangles (i.e. triplets of nodes with all three pairwise connections present) containing the node $i$ to the total number of node triplets constructed from the node $i$ and its neighbors (adjacent nodes) \cite{watts1998collective}. Equivalently, LCC is the fraction of directly connected (mutually adjacent) pairs of nodes among all pairs composed of the direct neighbors of the node $i$. According to \cite{watts1998collective}, LCC characterizes the ``cliquishness'' of the direct neighborhood of the node $i$, i.e. how close (in terms of the quantity of mutual connections) this neighborhood is to being a clique (which is a subgraph where all nodes are connected to each other). In particular, LCC equals one for any node which actually belongs to a clique composed of this node and all its neighbors (this is the case for the node $i$ in Fig.~\ref{fig_net_samples}$A$); conversely, if a node with all its neighbors constitute a star subgraph, then LCC of the star center is zero (this holds for the node $i$ in Fig.~\ref{fig_net_samples}$B$). Another instructive example is the Gilbert's formulation of Erd\H{o}s-R\'enyi random graphs \cite{gilbert1959random, erdos1959random}, where an edge between any pair of nodes is assumed to exist with the same preassigned probability $p$ independently on any other edges in the graph. In such random graphs, expected LCC of any node equals $p$, which becomes a small value in large networks with sparse connections; hence, the global clustering coefficient defined as average LCC over all nodes is also small, see discussion in \cite{watts1998collective}, \cite[Sec.~4.1]{newman2003structure}. In network science LCC along with the global clustering coefficient are commonly accepted as markers of so-called small-world networks, which are characterized, given the total number of nodes and edges, simultaneously by large values of clustering coefficients (as compared to Erd\H{o}s-R\'enyi random graphs) and small values of mean shortest path length (as compared to regular lattices); these features have been actually found in a variety of real networks \cite{watts1998collective, newman2003structure}.

Following the definition above, LCC may be expressed in terms of the adjacency matrix $(a_{ij})$ of an unweighted graph as
\begin{equation}\label{eq:LCC}
LCC_i^{\text{unw}} = \frac{\sum_{1\le j < l \le N (j,l\neq i)} a_{ij} \cdot a_{il} \cdot a_{jl}}
  {\sum_{1\le j < l \le N (j,l\neq i)} a_{ij} \cdot a_{il}}.
\end{equation}
Note that the denominator in \eqref{eq:LCC} is essentially the number of unordered pairs (2-combinations) among the nodes adjacent to $i$ and can be equivalently expressed as $k_i (k_i-1)/2$, where $k_i$ is the degree of the node $i$ (the number of the nodes adjacent to $i$). We do not use this replacement in the expression \eqref{eq:LCC} in order to facilitate establishing its correspondence to the weighted-network modifications of LCC formulation.

Multiple extensions of the LCC definition to account for link weights are available in the literature \cite{frontiers2018, barrat2004architecture, onnela2005intensity, zhang2005general, holme2007korean, opsahl2009clustering}. Among them, the most straightforward modification of Eq.~\eqref{eq:LCC} is the definition suggested in \cite{zhang2005general}, which, as shown in \cite{kalna2006clustering}, can be obtained from Eq.~\eqref{eq:LCC} simply by replacing all adjacency matrix elements $a_{ij}$ with the respective normalized link weights. Similar expressions were also derived in \cite{grindrod2002range, ahnert2007ensemble} based upon unweighted random graphs with probabilistic links, where the probability for a link to exist between a pair of nodes in a random graph is taken proportional to the respective link weight from the underlying deterministic weighted network. Comparative reviews of different formulations of LCC for weighted networks can be found in \cite{saramaki2007generalizations, wang2017comparison}.

The LCC formulation introduced in \cite{frontiers2018} is specifically tailored to correlation networks, which sets it apart from other LCC definitions adapted to weighted graphs in general. The approach of \cite{frontiers2018} implies that a correlation matrix is an empirical manifestation of some underlying interactions, and seeks to provide an empirical estimate of LCC adequate to the network of that underlying interactions. The idea behind the LCC definition in \cite{frontiers2018} is to compensate for the impact of indirect interaction paths (e.g. node $j$ interacting with node $i$, which in turn interacts with node $l$, as in both panels of Fig.~\ref{fig_net_samples}) upon the correlation matrix. Namely, the nodes $j$ and $l$ in the example above are expected to exhibit some level of correlation not only in the presence of an actual direct interaction between them (e.g. when the actual interactions are as in Fig.~\ref{fig_net_samples}$A$), but also even in the absence of such interaction (as in Fig.~\ref{fig_net_samples}$B$), due to indirect interaction paths (in particular, but not limited to, the path mediated by the node $i$). Consequently, such indirect correlations make the true underlying interaction structure hard to identify by the correlation matrix, and moreover, this may dramatically affect the empirical LCC value; e.g. the actual interaction structure shown in Fig.~\ref{fig_net_samples}$B$ may produce, in the worst case scenario, a correlation network with as many spurious connections as in Fig.~\ref{fig_net_samples}$A$, which translates into an inflation of LCC from 0 in the actual underlying interactions network to 1 in the empirical correlation network. In order to reduce the influence of such indirect correlations on the resultant LCC value, the definition of LCC in \cite{frontiers2018} makes use of the three-way partial Pearson's product-moment correlation coefficient $\rho^{\text{part}}_{jl|i}$ \cite{whittaker1990graphical}, which essentially indicates the surplus correlation between the nodes $j$ and $l$ beyond their indirect correlation through the node $i$.

In our study we modify the LCC definition from \cite{frontiers2018} by replacing all Pearson's product-moment correlation coefficients by the respective Kendall's rank correlation coefficients, which produces the expression
\begin{equation}\label{eq:LCCw}
LCC_i^{\text{wei}} = \frac{\sum_{1\le j < l \le N, (j,l\neq i)} | \tau_{ij} \cdot \tau_{il} \cdot \tau^{\text{part}}_{jl|i}|}{\sum_{1\le j < l \le N (j,l\neq i)} | \tau_{ij} \cdot \tau_{il}|}
\end{equation}
differing from the initial unweighted definition \eqref{eq:LCC} in the following: (i) the adjacency matrix elements $a_{ij}$ and $a_{il}$, which denote the connections between the node $i$ and its neighbors, are replaced by the respective Kendall's correlation coefficients $\tau_{ij}$ and $\tau_{il}$; (ii) the adjacency matrix element $a_{jl}$, denoting the mutual interaction between the node $i$'s neighbors $j$ and $l$, is replaced by the three-way partial Kendall's rank correlation coefficient $\tau^{\text{part}}_{jl|i}$, which in turn is defined according to \cite[Section~12.6]{gibbons2003nonparametric} as
\begin{equation}\label{eq:taupart}
\tau^{\text{part}}_{jl|i} = \frac{\tau_{jl} - \tau_{ij} \cdot \tau_{il} }{\sqrt{1 - \tau_{ij}^2} \sqrt{1 - \tau_{il}^2}};
\end{equation}
(iii) all correlation coefficients, including the partial one, are taken by absolute value, which implies that positive and negative correlations are treated equally. The latter is in line with the approach taken in \cite{frontiers2018}, which is substantiated by the reasoning that mutual correllations and partial correlations, regardless of their sign, are equally interpreted as indicators of interaction between the respective nodes. Notably, the expression \eqref{eq:taupart} is analogous to the definition of the three-way partial Pearson's correlation coefficient $\rho^{\text{part}}_{jl|i}$ \cite{frontiers2018, whittaker1990graphical} with all pairwise Pearson's correlations replaced by the respective Kendall's correlations. Despite this resemblance, the expression \eqref{eq:taupart} for $\tau^{\text{part}}_{jl|i}$ is not obtained merely by a formal substitution, but is properly justified based on the general concept of partial correlations, see details in \cite[Section~12.6]{gibbons2003nonparametric}.

As it is pointed out in \cite{frontiers2018}, the three-way partial correlation coefficient is intended to compensate for only one specific indirect interaction path between the nodes $j$ and $l$, namely, the path mediated by the node $i$ (for which LCC is being computed), and ignores other possible indirect interaction paths. This approach is well-suited to the triangle-based structure of the LCC definition; that said, taking into account \textit{every} indirect interaction path could arguably further improve the adequacy of the reconstructed network topology to the underlying interactions. However, this task is difficult in terms of both computational burden and poor tractability for statistical estimation \cite{brier2015partial}, the latter being especially important when the sample size (equal to the time window length) is rather small due to the imposed limitations, as discussed in Sec.~\ref{sec_evonet}. These difficulties render the mentioned improvement unfeasible.

As long as the expression \eqref{eq:taupart} contains a division by zero whenever $|\tau_{ij}=1|$ or $|\tau_{il}=1|$, such values of the indices $j$ and $l$ are excluded from summation both in the numerator and in the denominator of \eqref{eq:LCCw}, whenever this occurs (such cases turn out to be rare but not impossible in our actual computations). Essentially, as mentioned in Sec.~\ref{sec:kendall}, unity absolute value of Kendall's $\tau$ implies that the bivariate statistics within such a pair of variables (between the nodes $i$ and $j$, or $i$ and $l$) is indistinguishable by the data within the sample from a deterministic monotone functional dependence. We use this observation to justify the exclusion of such node pairs from summation, which may be interpreted as such a pair being temporarily (for a particular position of the time window) lumped together into a single node, as soon as they turn out to behave as a single node anyway. That said, excluding unity-correlated pairs of nodes inevitably introduces a methodical perturbation into the obtained LCC value, and the final justification for this operation relies mostly on our observation that such perturbations, being rare and relatively small, do not affect the capability of thus computed LCC as a climate marker.

We also note that although three-way partial correlations are still not expected to recover completely the true underlying interaction structure of whatever network system, we expect this approach, following the reasoning of \cite{frontiers2018}, to produce the most adequate and meaningful empirical estimate of LCC from correlation analysis of multivariate data among all available LCC formulations. In particular, it was found in \cite{frontiers2018} experimentally, that LCC defined in terms of three-way partial correlations better reveals the age dependence of brain connectivity by functional magnetic resonance data, as compared to other available LCC formulations for weighted networks \cite{barrat2004architecture, onnela2005intensity, zhang2005general}, and to the LCC definition \eqref{eq:LCC} for unweighted (thresholded) correlation networks. The present study, to the best of our knowledge, is the first to combine the three-way partial correlations approach of \cite{frontiers2018} with the Kendall's formulation of correlation coefficient in the definition of LCC for correlation networks \eqref{eq:LCCw}, and to apply this new LCC definition to climate networks.

As a local network measure, an LCC value is attributed to every network node. For an evolving network, LCC of all nodes are computed for each network instance, which can be represented by appending an additional temporal index $(m)$ to all variables in Eqs.~(\ref{eq:LCC}--\ref{eq:taupart}). The resultant LCC arrays thus constitute georeferenced multivariate time series $\{LCC_i^{\text{unw}}(t_m)\}$ and $\{LCC_i^{\text{wei}}(t_m)\}$ on the same spatial grid as the source data, and with the time grid of length $M$ (see Eq.~\eqref{eq_Ntime}) inherited from the evolving network.

\subsection{Empirical complementary cumulative distribution function transform}\label{sec_ecdf}

The nodes of the spatial grid are subject to unequal conditions due to the natural spatial inhomogeneity of the climate system (including the influence of land and all other objective reasons), and also due to the geometric non-uniformity of the grid itself (which is periodic in geographic coordinates, but not in actual space). This inhomogeneity in the source climate data is partially compensated by subtracting the respective local daily climate normals (see Sec.~\ref{sec_srcdata}). However, fluctuations and their inter-node correlation properties inevitably retain some inhomogeneity, thus affecting the constructed correlation networks and their graph-theoretic measures (here, different LCC formulations). As soon as the georeferenced time series of these measures (further referred to as indicators) are to be analysed for association with events of interest (TCs), which are also spatiotemporal phenomena, the indicator time series have to be transformed (regularized) such that their spatial inhomogeneity (across the network nodes) is compensated as far as possible prior to the association analysis. To this end, we use an approach based on the empirical complementary cumulative distribution function transform (a modification of rank transform), which we describe below in the general form, and apply it to the georeferenced LCC time series.

It is common in mathematical statistics to address the problem of non-identical, non-normal and a priori unknown distributions of variates by means of the rank transformation \cite{Conover1981}, on which a variety of non-parametric statistical methods are based \cite{gibbons2003nonparametric}, whereby the actual observations are replaced by their ranks, i.e. sequential numbers within a sorted sample. The rank transformation can be normalized by the sample size, so that the result fits into the $[0,1]$ range.

If the ranking order is \textit{ascending}, the normalized rank transformation coincides with the empirical cumulative distribution function (ECDF) \cite[Sec. 2.3]{gibbons2003nonparametric}. Using ECDF transformation to regularize input data is also a known practice in machine learning \cite{zhuang2020feature}. ECDF of a variate $x$ is defined as the empirical frequency of observing a value of $x$ which is \textit{same or smaller} than the function argument $\xi$ over the entire available sample (or time series) of the variate:
\begin{equation}\label{eq_ECDF}
ECDF_x(\xi) = \frac{1}{M} \sum_{m=1}^{M} 1\left(x^{(m)} \le \xi\right),
\end{equation}
where $M$ is the sample size (length of the time series), $x^{(m)}$ are the observations in the sample, notation $1(x^{(m)} \le \xi)$ has the same meaning as in Eq.~\eqref{eq_thr}.

Under certain conditions (see below) $ECDF_x(\xi)$ is an unbiased consistent empirical estimate of the underlying cumulative distribution function $CDF_x(\xi)$, which means that $ECDF_x(\xi)$ has expectation equal to $CDF_x(\xi)$, and converges to it (e.g. in the mean-square sense, meaning that variance tends to zero) when the sample size increases \cite[p.~18]{gibbons2003nonparametric}. In turn, CDF is known to transform any continuous random variate so that the outcome is uniformly distributed on $[0,1]$; precisely, the transformed variate $y=CDF_x(x)$ is distributed uniformly on $[0,1]$, whenever $CDF_x(\cdot)$ is any continuous function \cite[Sec.~2.5]{gibbons2003nonparametric}, \cite[Proposition 2]{embrechts2013note}. This substantiates the practice of using ECDF (as an empirical estimate of CDF) to compensate for non-identical and unknown distributions in multivariate data by reducing all variates to a known universal distribution (namely, uniform on $[0,1]$) in the limit of large sample size.

In turn, the \textit{descending} ranking order may be considered similarly. In this case, the normalized rank transformation is obtained by reversing the inequality sign in the expression \eqref{eq_ECDF}. This produces a function that we refer to as the empirical complementary cumulative distribution function 
\begin{equation}\label{eq_ECCDF}
ECCDF_x(\xi) = \frac{1}{M} \sum_{m=1}^{M} 1\left(x^{(m)} \ge \xi\right)
\end{equation}
with same notations as in Eq.~\eqref{eq_ECDF}. Note that $ECCDF_x(\xi)$ defined this way does not actually complement $ECDF_x(\xi)$ to unity ($ECCDF_x(\xi)+ECDF_x(\xi) \not\equiv 1$); however, given a continuous variate $x$, $ECCDF_x(\xi)$ provides an unbiased consistent estimate of the underlying complementary cumulative distribution function
\begin{equation}
CCDF_x(\xi)=1-CDF_x(\xi),
\end{equation}
and, similarly to a mentioned property of CDF, the transformed variate $y=CCDF_x(x)$ is distributed uniformly on $[0,1]$ under the same assumptions.

In order to choose between the ECDF and ECCDF transformations, we note that a natural measure to quantify an observed anomaly of a georeferenced variable is the empirical frequency of observing a \textit{same or stronger} anomaly in the same location over the entire observation time, so that the closer is this frequency to zero, the greater the anomaly. When anomalies of interest are \textit{small values} (so that a stronger anomaly implies a smaller value), the abovementioned frequency of same or stronger anomalies coincides with the definition of ECDF \eqref{eq_ECDF}. Conversely, when anomalies of interest are \textit{high values}, they are to be characterized by the empirical frequency of observing \textit{same or greater} values, which coincides with ECCDF \eqref{eq_ECCDF}. As soon as TCs turn out to be associated with \textit{anomalously high} values of LCC \cite{gupta2021complex}, we use the ECCDF transformation \eqref{eq_ECCDF}. Due to the argument above, this transformation not only compensates for the non-identical and apriori unknown distributions within multivariate time series, but is also meaningful as the frequency of same or stronger anomalies for the particular location. 

The required property of ECDF (equivalently, ECCDF) to be an unbiased consistent estimate of CDF (CCDF) is typically proven under the assumptions of independence and identical distribution of the observations in the sample \cite[Sec.~2.3]{gibbons2003nonparametric}. However, if the sample is constructed from a time series, then typically one or both assumptions are not fulfilled; in this case, the time series $\{x^{(m)}\}_{m=1..M}$ have to be considered as a realization of an underlying stochastic process $\{\tilde{x}^{(m)}\}$ \cite[Chap.~9]{papoulis2002probability}, which is a sequence of (generally, dependent) random variates numbered by the temporal index $m$. For consistency with cited literature, in the reasoning below we speak of ECDF and CDF; however, the same applies equivalently to ECCDF and CCDF. Generally speaking, the variates $\tilde{x}^{(m)}$ may be differently distributed; in this case there is no single distribution $CDF_x(\cdot)$ describing all observations. This problem is eliminated if the stochastic process is \textit{first-order stationary} \cite[p.~392]{papoulis2002probability} (i.e. stationary with respect to distribution of an individual observation), meaning that the distribution functions $CDF_{\tilde{x}^{(m)}}(\cdot)$ of all individual variates $\tilde{x}^{(m)}$ are identical for all $m$, and thus they admit a single notation $CDF_{x}(\cdot)$ without a temporal index (see discussion below on the applicability of this assumption in the context of the present study). If this holds, then the expectation of each term under summation in Eq.~\eqref{eq_ECDF} is
\begin{equation}\label{eq_unbias}
\left<1\left(x^{(m)} \le \xi\right)\right>=P\left(x^{(m)} \le \xi\right)=CDF_x(\xi)
\end{equation}
for any $m$, and thus the expectation $\left<ECDF_x(\xi)\right>$ always equals $CDF_x(\xi)$, or, in other words, $ECDF_x(\xi)$ is an unbiased estimate of $CDF_x(\xi)$. In turn, if the underlying distributions $CDF_{\tilde{x}^{(m)}}(\cdot)$ are not identical, then the expectation of $ECDF_x(\xi)$ equals the arithmetic mean of these distribution functions over the time grid.

Furthermore, the underlying process must be \textit{distribution-ergodic} \cite[p.~536]{papoulis2002probability}  (or ergodic with respect to distribution of an individual observation), which means that the estimate \eqref{eq_ECDF} converges to the limit (e.g. in the mean-square sense) when the sample size $M$ increases. This is equivalent to the definition of a consistent estimate \cite[p.~19]{gibbons2003nonparametric}. A sufficient condition for a stationary process $\{\tilde{x}^{(m)}\}$ to be distribution-ergodic (or, equivalently, for $ECDF_x(\xi)$ to be a consistent estimate of $CDF_x(\xi)$) is the assumption that a pair of individual variates $\tilde{x}^{(m)}$ and $\tilde{x}^{(m^{\prime})}$ become independent, or their dependence vanishes quickly enough, whenever they are separated by a sufficiently large time interval, i.e. when $|m-m^{\prime}|$ is large enough \cite[p.~537]{papoulis2002probability}. In general, the convergence of a time mean (such as in Eq.~\eqref{eq_ECDF})) to the respective underlying expectation is essentially a consequence of \textit{ergodic theorems}, which exist in a variety of formulations, e.g. pointwise (Birkhoff's) for almost sure convergence \cite{birkhoff1931proof}, or mean (von Neumann's) for mean-square convergence \cite{neumann1932proof,birkhoff1939mean}. Note that a stochastic process is not necessarily required to be stationary in order to be ergodic. The ergodicity concept has been extended to a more general class of \textit{asymptotically stationary} processes, which includes periodic nonstationarities (see \cite{boyles1983cycloergodic,anh1991covariance} and references therein), so that if an asymptotically stationary process is ergodic, then an arithmetic mean over a finite realization (time series) converges with increased realization length to a meaningful limit, namely, to the pooled (time-averaged) expectation \cite{BHAGAVAN1985311}, despite the nonstationarity. To summarize, when the requirements of stationarity (or asymptotic stationarity) and ergodicity of the underlying process are met, a finite but sufficiently large time series $\{x^{(m)}\}_{m=1..M}$ can be used as a sample of the variate $x$ in Eqs.~\eqref{eq_ECDF} and \eqref{eq_ECCDF}, despite a possible mutual dependence of observations within the series; however, if nonstationarity (e.g. periodic one) is admitted, then its effect upon the result has to be addressed in subsequent analysis.

Physically, the assumption of stationarity means that any changes in overall conditions during the observation time are neglected, or that a possible effect of such changes is not in the scope of the study. Climate data are actually subject to nonstationarity, which includes daily and annual cycles, and long-term trends of climate change. We do not expect daily cycle to manifest in the LCC time series, as soon as the time window used to construct evolving correlation networks (see Sec.~\ref{sec_evonet}) is a multiple of 24 hours, and, thus, always contains an integer number of daily periods. We also do not expect the relatively small climate change during the available observation time (which is almost 4 decades, see Sec.~\ref{sec_srcdata}) to impact significantly the phenomena under study (i.e. MSLP correlation network indicators and their association with TCs). At the same time, a possible annual nonstationarity in the LCC time series may influence the results of our analysis, especially in view of considerable seasonal variation in the frequency of tropical cyclones; this issue is addressed in Sec.~\ref{sec_assoc}.

In turn, ergodicity of time series generally arises from the underlying dynamics, as described by the ergodic theory  \cite{cornfeld2012ergodic,coudene2016ergodic,Wiggins2004Foundation}.
Climate dynamics is commonly assumed to be ergodic, when climate change can be neglected; for discussion of the interplay between climate change and ergodicity, see \cite{tel2020theory}. As discussed above, periodic (including annual) nonstationarity does not hinder ergodicity, and all other nonstationarities are neglected; thus, we treat the time series under study as realizations of ergodic processes.

In order to speak of ECDF (ECCDF) as an approximation of the underlying CDF (CCDF), it is also necessary to assume that the sample size is sufficient, that is, the observation time interval covered by the available (finite) time series encompasses a sufficient number of effectively independent observations (despite their dependence within short time intervals). This is provided by the observation time greatly exceeding both the annual period and the relevant correlation time scales in the dynamics of the climate system.

We justify all the assumptions made above by comparing our final results to a reference obtained from surrogate data (see Sec.~\ref{sec_assoc}, Sec.~\ref{sec_Results}).

Note that using the rank transform or, equivalently, its normalized modifications (ECDF or ECCDF) for regularizing differently distributed variates in multivariate data implies that the transform is computed individually for each variate (independently on other variates) \cite{Conover1981, zhuang2020feature}. In our case this implies that ECCDF transform is computed \textit{locally} for each individual spatial grid node, independently on other nodes, based on the entire available time series of the indicator under study for this specific node (i.e. for the particular point on the Earth surface).

The considerations above lead to the following procedure of multivariate time series transformation (regularization). \textit{General input data} are time series denoted below as $\{x_i^{(m)}\}$. The \textit{actual input data} are the georeferenced time series of the two LCC formulations $\{LCC_i^{\text{unw}}(t_m)\}$ and $\{LCC_i^{\text{wei}}(t_m)\}$ obtained from the procedure described in Sec.~\ref{sec_LCC}; in actual computations, they are substituted in turn for $\{x_i^{(m)}\}$ in the expressions below.

We consider each individual component of the multivariate time series $\{x_i^{(m)}\}$ (at any given $i$) as a statistical sample of a random variate $x_i$ in the sense which follows from the assumptions of (at least, asymptotic) stationarity and ergodicity, as discussed above. Then, writing down the ECCDF transform \eqref{eq_ECCDF} for each variate $x_i$ instead of the generic variate $x$, and substituting each individual observation $x_i^{(m)}$ in turn for the function argument $\xi$, we express the transformation as
\begin{multline}\label{eq_ECCDFmulti}
f_{i}^{(m)} = ECCDF_{x_i}\left(x_i^{(m)}\right) =\\ \frac{1}{M} \sum_{m^{\prime}=1}^{M} 1\left(x_i^{(m^{\prime})} \ge x_i^{(m)} \right),
\end{multline}
where the result of the transformation constitutes new multivariate time series of the transformed indicator $\{f_{i}^{(m)}\}$.

Under the assumptions of stationarity and ergodicity of the underlying stochastic process for any $i$, the transformation \eqref{eq_ECCDFmulti} converges (in the mean-square sense) at large sample size $M$ to the complementary CDF
\begin{equation}\label{eq_toCCDF}
f_{i}^{(m)} \xrightarrow[M\to\infty]{} 1 - CDF_{x_i}(x_i^{(m)}),
\end{equation}
so that each $f_{i}^{(m)}$ can be seen as an observation of a new (transformed) variate $f_i=CCDF_{x_i}(x_i)$, which is distributed uniformly on $[0,1]$ for any $i$ by construction. This does not hold if the stationarity assumption is replaced by asymptotic stationarity; in this case, the time mean in Eq.~\eqref{eq_ECCDFmulti} converges to the pooled (i.e. time-averaged \cite{BHAGAVAN1985311}) CCDF
\begin{equation}\label{eq_toCCDFnonst}
f_{i}^{(m)} \xrightarrow[M\to\infty]{} 1 - \overline{CDF}_{x_i}(x_i^{(m)}),
\end{equation}
where
\begin{equation}
\overline{CDF}_{x_i}(\xi) = \lim_{M\to\infty} \frac{1}{M} \sum_{m=1}^M CDF_{\tilde{x}_i^{(m)}}(\xi).
\end{equation}
However, if the nonstationarity is periodic and small (which is assumed), then each individual distribution $CDF_{\tilde{x}_i^{(m)}}(\xi)$ weakly differs from the pooled one, thus the distributions of the transformed variates remain close to the uniform one, although having a small modulation in time with the period of the nonstationarity.

Note an instructive case when all observations in the sample $\{x_{i}^{(m)}\}_{i=const,m=1..M}$ are different (there are no ties, i.e. no such $m$ and $n$ that $m\neq n$ and $x_i^{(m)}=x_i^{(n)}$); then, the sample of any variate $f_i$, i.e. the set $\{f_{i}^{(m)}\}_{i=const,m=1..M}$ computed according to Eq.~\eqref{eq_ECCDFmulti}, is automatically (without any further assumptions) a permutation of the equidistant grid $\{k/M\}_{k=1..M}$. 

Furthermore, according to Eq.~\eqref{eq_ECCDFmulti}, each individual transformed observation $f_{i}^{(m)}$ equals the frequency of observations $x_i^{(m^{\prime})}$ satisfying the condition $x_i^{(m^{\prime})} \ge x_i^{(m)}$ within the sample $\{x_i^{(m^{\prime})}\}_{i=const,m^{\prime}=1..M}$, i.e. the empirical frequency of observing a value of $x_i$ greater than or equal to the respective observation $x_{i}^{(m)}$ at given $i$.

When the georeferenced time series of a chosen indicator ($\{LCC_i^{\text{unw}}(t_m)\}$ or $\{LCC_i^{\text{wei}}(t_m)\}$) are substituted instead of $\{x_i^{(m)}\}$ in the ECCDF transformation \eqref{eq_ECCDFmulti}, any particular (``current'') observation within the transformed indicator time series $f_{i}^{(m)}$ (at given $i$ and $m$) means the empirical frequency of observing a same or stronger anomaly of the indicator in the particular point on the Earth surface (at the node $i$) over the entire observation time, as compared to the current (at the same $i$ and $m$) observation of this indicator. Note that if the described methodology is applied to indicators expected to demonstrate anomalously low values associated to the event of interest, then, for the interpretation above to remain valid, the ECDF transformation \eqref{eq_ECDF} should be used instead of ECCDF; this results in reversing the inequality sign in Eq.~\eqref{eq_ECCDFmulti}.

\subsection{Nonlocal association of georeferenced indicator time series to TC trajectories}\label{sec_assoc}
Our goal is to characterize and compare the extent to which the indicators under study (here, different LCC formulations) demonstrate anomalous behavior associated with events of interest (TCs). While indicators are represented by georeferenced time series, TCs are represented by Best Track data \cite{mohapatra2012best}, which consist of TC center trajectories and TC strength information over time. Since we currently ignore the TC strength information and use only TC trajectories, the problem can be formulated in terms of establishing association between two different types of spatio-temporal data: georeferenced transformed indicator time series $\{f_{i}^{(m)}\}$ (obtained from the procedure described in Sec.~\ref{sec_ecdf}) on the one hand, and a set of TC trajectories on the other. Due to the absence of apriori information on the spatial positioning of a TC-associated indicator anomaly relative to the TC center (in particular, an indicator anomaly in general may even preferably occur not at the TC center, but at a distance), the mentioned association problem is nonlocal in space, in the sense that we speak of association whenever an indicator anomaly and a TC are observed at the same time within a certain (apriori unknown) allowable distance in space.

We quantify this association by defining a spatial vicinity of a TC center in any moment of the time grid, introducing the null hypothesis of independence between the indicator and TC trajectories, and posing the question whether (and to what extent) the actual (empirical) statistics of indicator observations taken within the joint vicinity of all TCs (i.e. conditionally upon belonging to a TC center vicinity) over the entire time of observation differs from the result expected under the null hypothesis.
Namely, under this null hypothesis and with stationarity assumed, the conditional distribution of $f_{i}^{(m)}$ must coincide with the unconditional, which, in turn, is uniform on $[0,1]$ by construction (see Sec.~\ref{sec_ecdf}). Therefore, the divergence of the empirical conditional distribution of $f_{i}^{(m)}$ from the uniform one will be interpreted as a measure of association (dependence) between the indicator under study and the condition (of belonging to a TC center vicinity); at the same time, the effect of possible annual nonstationarity of the indicator time series has to be assessed. This reasoning is formalized and detailed below. 

The size (and the shape) of the mentioned spatial vicinity is a free parameter of the method. For the sake of simplicity, we define the vicinity of a TC at a given time $t_m$ as a quadratic-shaped subarray of the spatial grid of a specified size $R$, centered closest to the current position of the TC center. This subarray is denoted as $v_c^{(m)}(R)$, where $c$ denotes a TC from the available set. We tried subarray sizes ranging from $2\times 2$ grid points ($R=1$) up to $12\times 12$ points ($R=6$). Note that the spatial grid is uniformly spaced in terms of geographic coordinates, but not in terms of their actual distance on the Earth surface, because equal spacing in longitude translates into different spacings in distance when taken at different latitudes. The actual spacing of the grid points is proportional to the cosine of the latitude, which in the region of interest (5\textdegree N to 30\textdegree N latitude) produces the greatest difference of approximately 15\%, which we neglect for simplicity, although this effect may need to be taken into account in the polar regions.

To each space-time grid point $(i,m)$ we ascribe a binary (taking on values `yes' or `no') attribute $V_{R,i}^{(m)}$, which answers the question whether this grid point belongs to a vicinity of size $R$ of some TC center at the respective time $t_m$, according to the following rule:
\begin{equation}\label{eq_Vim}
V_{R,i}^{(m)}= \begin{cases}
\text{yes},\quad \text{if } \exists c \text{ so that } i\in v_c^{(m)}(R) ,\\
\text{no}, \quad \text{otherwise}.
\end{cases}
\end{equation}

Consider a subset $S_{V}$ of the space-time grid $\{(i,m)\}$ satisfying the condition $V_{R,i}^{(m)} = \text{yes}$:
\begin{equation}\label{eq_Sv}
S_{V}=\{(i,m)| V_{R,i}^{(m)} = \text{yes}\},
\end{equation}
which is the joint vicinity of all TC trajectories. Then, consider a conditional ECDF defined similarly to Eq.~\eqref{eq_ECDF} over a sample of observations $\{f_{i}^{(m)}\}_{(i,m)\in S_V}$, which are taken conditionally upon the spacetime position of the observation belonging to $S_{V}$, i.e. upon $V_{R,i}^{(m)} = \text{yes}$, as follows:
\begin{equation}\label{eq_ECDFcond}
ECDF_{f|S_{V}}(\xi) =\frac{1}{|S_{V}|} \sum_{(i,m)\in S_{V}} 1\left(f_{i}^{(m)}\le\xi\right),
\end{equation}
where $|S_{V}|$ is the cardinality (number of elements) of the set $S_{V}$.
Following the lines of Sec.~\ref{sec_ecdf}, we consider the multivariate time series $\{f_{i}^{(m)}\}$ and $\{V_{R,i}^{(m)}\}$ as realizations of underlying multivariate stochastic processes $\{\tilde{f}_{i}^{(m)}\}$ and $\{\tilde{V}_{R,i}^{(m)}\}$. As soon as the index $i$ maps into space (i.e. on the Earth surface), these processes are essentially random fields, which means a random function of point in spacetime (here, discrete).

If the process (field) $\{\tilde{f}_{i}^{(m)}\}$ is stationary (which follows from $\{\tilde{x}_i^{(m)}\}$ being stationary), then all individual variates $\tilde{f}_{i}^{(m)}=CDF_{x_i}(\tilde{x}_{i}^{(m)})$ are identically distributed (namely, uniformly on $[0,1]$) in the limit of large $M$ by construction (see Sec.~\ref{sec_ecdf}); hence (similarly to Eq.~\eqref{eq_unbias}), Eq.~\eqref{eq_ECDFcond} is an unbiased estimate of the underlying conditional CDF. Similarly to the argument in Sec.~\ref{sec_ecdf}, the assumption of distribution-ergodicity of the underlying random field implies that the arithmetic mean in Eq.~\eqref{eq_ECDFcond} converges (e.g. in the mean square) with increasing sample size $|S_{V}|$ to the conditional expectation of the quantity being averaged, and finally reduces to the conditional CDF, thus providing its consistent estimate:
\begin{multline}\label{eq_toCDFcond}
ECDF_{f|S_V}(\xi) \xrightarrow[|S_{V}|\to\infty]{} \left\langle 1\left(f_{i}^{(m)}\le\xi\right) \big| (i,m)\in S_V \right\rangle\\ = P\left(f_{i}^{(m)}\le\xi \big|  V_{R,i}^{(m)}=\text{yes} \right) \\= CDF_{f|V}\left(\xi \big|  V_{R,i}^{(m)}=\text{yes}\right).
\end{multline}
Moreover, under the null hypothesis, the random fields $\{\tilde{f}_{i}^{(m)}\}$ and $\{\tilde{V}_{R,i}^{(m)}\}$ are independent, and, thus, the conditional distribution equals the unconditional one
\begin{equation}
CDF_{f|V}\left(\xi \big|  V_{R,i}^{(m)}=\text{yes}\right) = CDF_{f}\left(\xi \right),
\end{equation}
which is uniform on $[0,1]$ by construction under the assumptions taken.

A difference of the argument above as compared to the reasoning behind Eq.~\eqref{eq_toCCDF} in Sec.~\ref{sec_ecdf} is that the arithmetic mean in Eq.~\eqref{eq_ECDFcond} is taken over a subset $S_V$ of the (3-dimensional) space-time grid, in contrast to a mean over a 1-dimensional time grid in Eq.~\eqref{eq_ECDF}. It implies that the distribution-ergodicity condition found in \cite[p.~536]{papoulis2002probability}, which was used in Sec.~\ref{sec_ecdf}, can not be applied here directly, because the consideration in \cite{papoulis2002probability} is essentially limited to the case where stochastic process realizations are averaged over a single dimension (time). Applying the concept of ergodicity to justify convergence of an arithmetic mean of a random field over a multi-dimensional space requires an accordingly generalized ergodic theorem. Such a generalization was first obtained in \cite{Wiener1939} for a continuous space; alternative formulations and proofs, and also further generalizations can be found e.g. in \cite{pitt_1942,Becker1981,anh1991covariance}; in particular, a generalization for discrete multidimensional spaces (i.e. for random fields defined over a multidimensional grid, which is the case in Eq.~\eqref{eq_ECDFcond}) is available in \cite[Sec.~4]{pitt_1942}, and for nonstationary (asymptotically stationary) fields (see Sec.~\ref{sec_ecdf}) in \cite{anh1991covariance}. Physically, a sufficient condition to justify the limit transition in Eq.~\eqref{eq_toCDFcond} may be formulated similarly to the one-dimensional case (cf. Sec.~\ref{sec_ecdf}) as the requirement that a pair of variates $\tilde{f}_{i}^{(m)}$ and $\tilde{f}_{i^{\prime}}^{(m^{\prime})}$ become independent, or their dependence decays quickly enough, when the distance between the points of the space-time grid $(i,m)$ and $(i^{\prime},m^{\prime})$ is sufficiently large.

Also similarly to the one-dimensional case (cf. Sec.~\ref{sec_ecdf}), the finite-sample empirical conditional distribution \eqref{eq_ECDFcond} is expected to approximate well its infinite-sample limit \eqref{eq_toCDFcond},
if the set $S_V$ is large enough that the (conditional) sample $\{f_i^{(m)}\}_{(i,m)\in S_V}$ encompasses a sufficient number of effectively independent observations, despite their dependence within short time intervals and short spatial distances. The physical justification of this assumption relies on the consideration that the set $S_V$, which is the joint vicinity of all TCs, covers a large number of TCs scattered over the entire observation time, which greatly exceeds the relevant correlation time scales of the climate dynamics.

In turn, in case of periodic nonstationarity, the distributions of the variates $\tilde{f}_{i}^{(m)}$ vary in time in the vicinity of the uniform distribution (see Sec.~\ref{sec_ecdf}) with the period of the nonstationarity. The ergodic theorem for asymptotically stationary (which include periodic) random fields \cite[Theorem 2]{anh1991covariance} requires that the subset over which the mean is taken (here, $S_V$) is rectangular and grows infinitely in all dimensions. Physically, when the subset encompasses a sufficient number of nonstationarity periods, the nonstationarity effectively averages out, thus ensuring ergodicity (convergence of the mean). Note, however, that the actual set $S_V$ in Eq.~\eqref{eq_ECDFcond} does not satisfy this requirement, because essentially it is a union of disjoint TC vicinities in spacetime, while the duration of each TC is much shorter than the annual cycle, and the frequency of TC occurrence is also subject to annual nonstationarity. Together with annual nonstationarity of the indicator time series, this hinders the ergodic limit transition in Eq.~\eqref{eq_toCDFcond}, making the outcome of the averaging in Eq.~\eqref{eq_ECDFcond} dependent on the properties of the set $S_V$. For example, as TCs in the region of interest (Northern Indian Ocean) are known to occur most frequently during the SOND season (September through December) \cite{Sugi2002}, spacetime grid nodes falling within this season will be prevalently present in $S_V$. This is not an issue with stationary indicator time series (because then all moments of time are equivalent), but an annual nonstationarity (e.g. if an indicator by itself is prone to anomalies within the SOND season, regardless of TCs) makes the average over $S_V$ in Eq.~\eqref{eq_ECDFcond} differ from the global (unconditional) average even under the null hypothesis. 

In order to assess the magnitude of the influence of annual nonstationarity upon our results, we initially perform computations assuming the LCC time series to be stationary, and then repeat the processing stages described in Sec.~\ref{sec_ecdf}--\ref{sec_assoc} on a subset of the LCC time series consisting only of observations falling within the SOND season of each year, instead of the full time series. The deviation between the results obtained from the SOND-only and full time series are interpreted as a measure of the effect of annual nonstationarity (see Sec.~\ref{sec_Results}).
 
In turn, a deviation of the empirical conditional distribution \eqref{eq_ECDFcond} from the uniform one is an empirical evidence in support of rejecting the null hypothesis, i.e. an evidence of association (dependence) between the indicator and TC trajectories.

For the sake of visualization, the mentioned empirical distributions are represented in the form of probability density plots, instead of cumulative distributions \eqref{eq_ECDFcond}. The empirical conditional probability density functions $w_{f|S_V}(\xi)$ are obtained from the conditional sample $\{f_{i}^{(m)}\}_{(i,m)\in S_V}$ using the kernel density estimation method \cite{rosenblatt1956remarks,parzen1962estimation} implemented in the Scikit-Learn software library \cite{scikit-learn}.

In order to ensure the correctness of all assumptions, we test the above methodology against surrogate data, which are obtained by shifting the TC trajectories from the actual TC Best Track data by a random amount along the time grid. This shift simulates the fulfillment of the null hypothesis, i.e. independence between indicators under study and TC tracks. The time series $\{{V}_{R,i}^{(m)}\}$ (and, hence, the conditional sample $\{f_{i}^{(m)}\}_{(i,m)\in S_V}$) are constructed again based on these surrogate data according to Eqs.~\eqref{eq_Vim}, \eqref{eq_Sv}, and conditional empirical probability densities calculated based on this (surrogate) conditional sample, all other data same as in the main computation.

\section{Results}\label{sec_Results}
We have processed the georeferenced time series of MSLP anomaly (see Sec.~\ref{sec_srcdata}) by applying sequentially the procedures described in Sec.~\ref{sec_evonet}--\ref{sec_assoc}, as follows. The weighted evolving correlation network was constructed by computing all pairwise Kendall's correlation coefficients in a sliding time window over the source time series, as described in Sec.~\ref{sec_evonet}; the unweighted evolving networks were obtained by thresholding the correlation coefficients, so that the edge density (the proportion of connected node pairs among the total combinatorial number of node pairs) is a specified value. We constructed unweighted networks with edge density taken equal to 5\% and 10\%. Time series of local clustering coefficients were calculated for the weighted and unweighted evolving networks using the respective formulations in Sec.~\ref{sec_LCC}. To provide an assessment of the impact of annual nonstationarity in the LCC time series upon the final result, additional LCC time series (further referred to as SOND-only data) were prepared by keeping only the data for September through December of each calendar year and omitting the rest. The ECCDF transform, as given by Eq.~\eqref{eq_ECCDFmulti}, was applied to each of the available LCC time series (to the full and SOND-only data, for the weighted and the unweighted networks). Based on the transformed indicator time series and Best Track data for the trajectories of tropical cyclone centers (see Sec.~\ref{sec_srcdata}), empirical conditional probability density functions $w_{f|S_V}(\xi)$ (i.e. empirical density of ECCDF-transformed indicator values observed within the joint set of all TC vicinities) were computed as described in Sec.~\ref{sec_assoc}. Computations were performed with the sizes of the quadratic-shaped spatial vicinity $v_c^{(m)}(R)$ varied from $2\times 2$ to $12\times 12$ spatial grid nodes. Surrogate data for TC trajectories were obtained by shifting the actual trajectories in time, so that independence between TCs and LCC time series is simulated, and the same empirical conditional probability density functions were calculated also with these surrogate data.

Recall that the transformed indicator value by itself means the unconditional frequency of same or stronger anomalies of the indicator for the particular location, and is distributed uniformly on $[0,1]$ with the unconditional probability density equal to unity (see Sec.~\ref{sec_ecdf}). Hence, the value of the conditional probability density $w_{f|S_V}(\xi)$ (e.g. the hight of a peak) denotes how much more frequently indicator anomalies with given unconditional probability $\xi$ are observed in the vicinity of a TC, than unconditionally. We compare plots of the calculated empirical conditional distributions visually, and interpret a greater deviation of the conditional distribution from the uniform one as a stronger dependence between the particular indicator and the events of interest (TCs).

\begin{figure*}
	\centering
	(a)\includegraphics[width=0.47\textwidth]{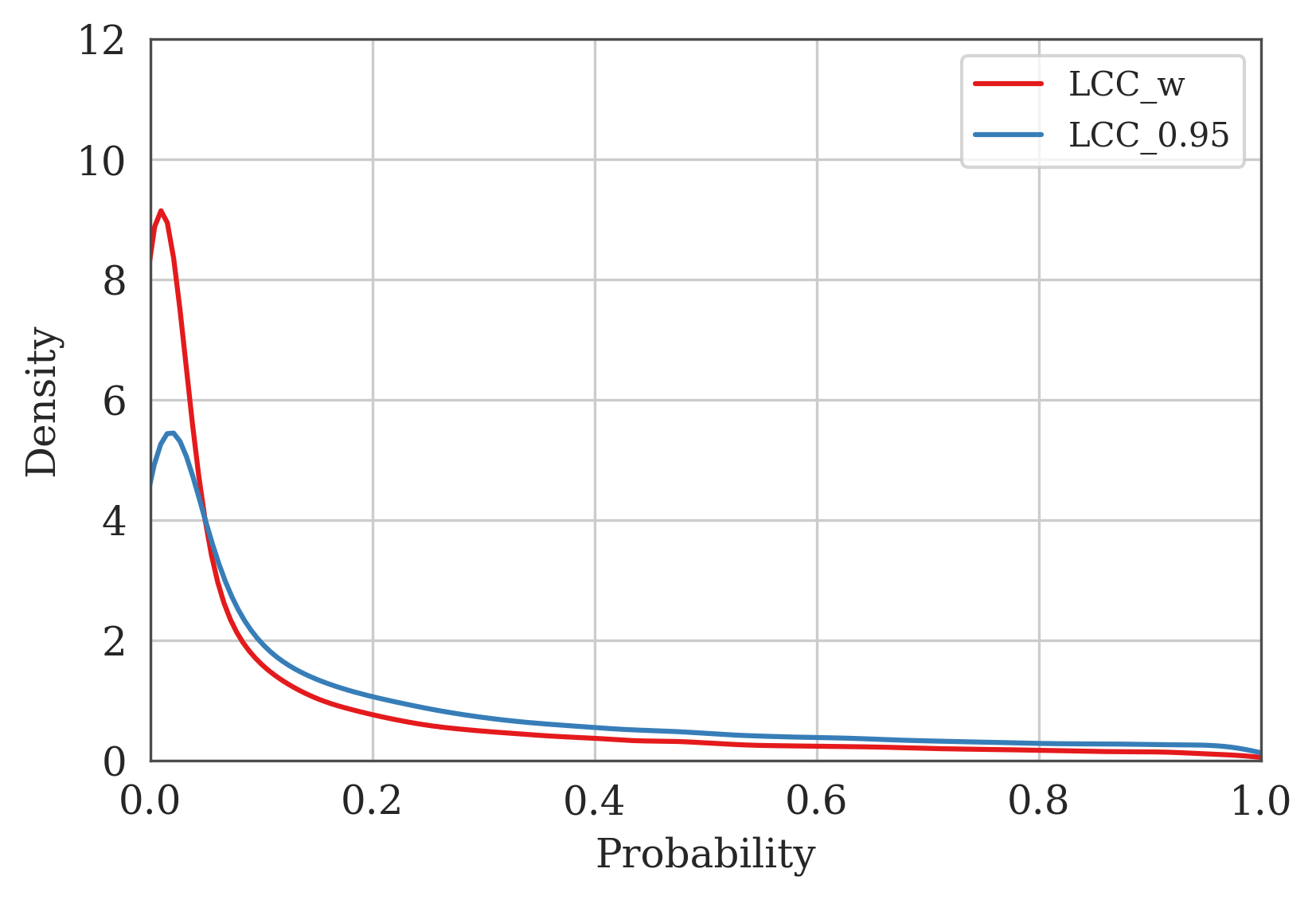}\hfill
	(b)\includegraphics[width=0.47\textwidth]{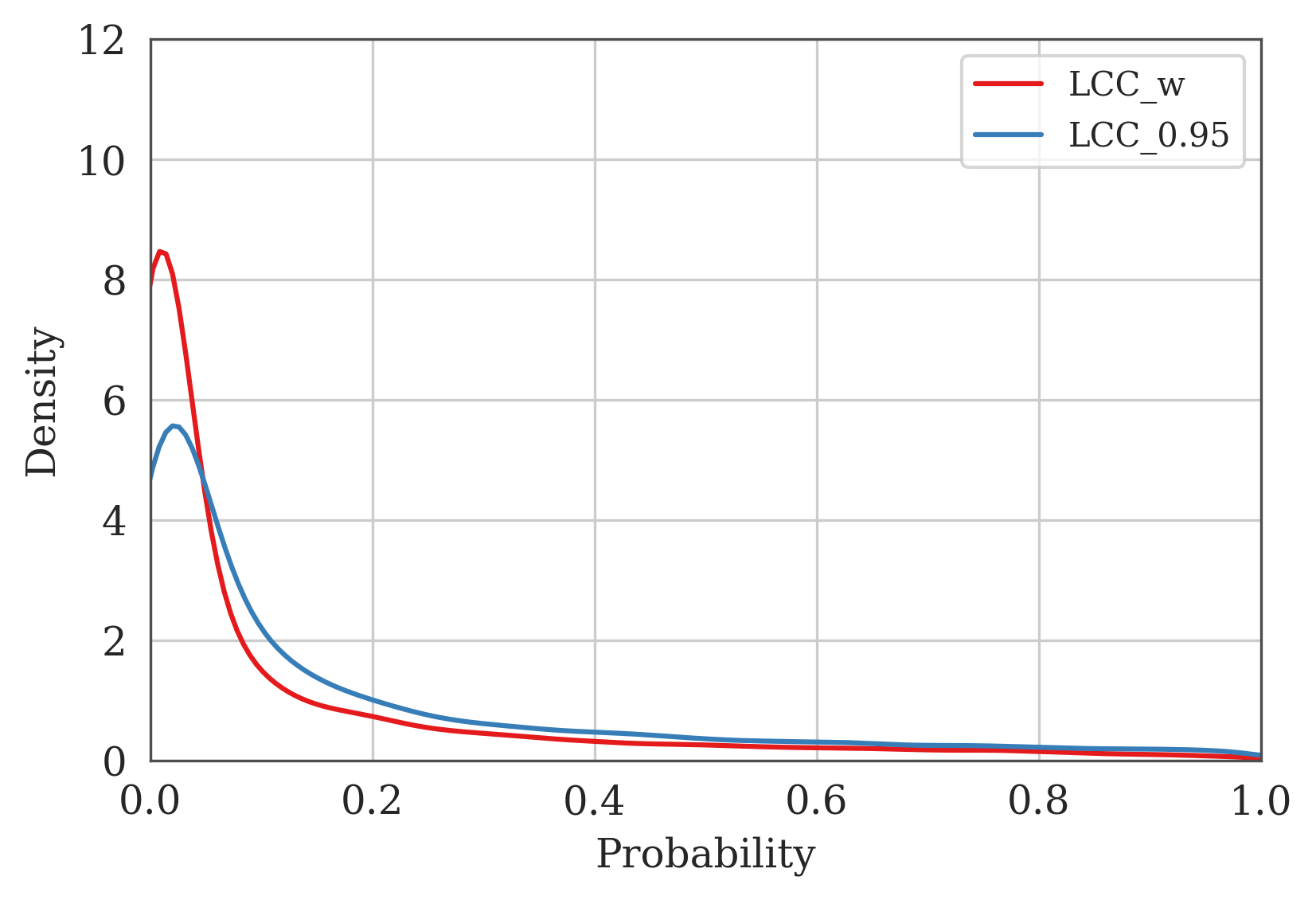}
	\caption{Empirical conditional probability density plots constructed from ECCDF-transformed indicator observations taken within the joint vicinity of all TC trajectories. Abscissa axis: value of a transformed indicator, ordinate axis: probability density. Indicators are local clustering coefficients for the weighted (red lines) and unweighted (blue lines, edge density is 10\%) correlation networks. Vicinity size is $4\times 4$ spatial grid nodes in panel (a) and $2\times 2$ in panel (b). Solid (dashed) lines denote the results obtained from the full (SOND-only) data, and dash-dotted lines --- from surrogate data.}\label{fig_mainres}
\end{figure*}

Examples of the obtained empirical conditional probability density functions are shown in Fig.~\ref{fig_mainres}. The results for the unweighted networks with edge density equal to 5\% and 10\% do not differ much; the latter demonstrate slightly better association with TCs and hence are shown in the plots. The strongest association (i.e. the greatest deviation of the empirical conditional distribution from the uniform one) for the weighted LCC is obtained for spatial vicinity of size $4\times 4$ spatial grid nodes (panel (a) in Fig.~\ref{fig_mainres}), and for the unweighted LCC for spatial vicinity of size $2\times 2$ nodes (panel (b) in Fig.~\ref{fig_mainres}). The weighted LCC formulation shows stronger association with TCs than the unweighted one, as quantified by an almost twice higher peak in the conditional probability density, which means an almost twice more frequent occurrence of the respective (having equal unconditional frequency) anomalies in the TC vicinity for the weighted as compared to the unweighted LCC formulation.

\section{Conclusion}
We have developed a broadly applicable statistical methodology, which is based upon normalized rank transformation and conditional empirical distributions, aimed at studying association (dependency) between an indicator represented by georeferenced multivariate time series and a class of ``events of interest'', which may be represented as an ensemble of spatiotemporal trajectories, or in any form reducible to a binary-valued (taking on values `yes' or `no') time series over the same space-time grid. The indicator time series are interpreted as a realization of a stochastic process (or random field), which is assumed to be stationary and ergodic with respect to probability distribution of an individual observation. Stationarity means that this distribution does not depend on the observation time, and ergodicity is provided when pairwise dependence between observations taken in different points of spacetime decays quickly enough (i.e. the pairwise observation distribution factorizes into a product of individual observation distributions) with increased distance in spacetime between these observations. In turn, the indicator probability distribution is not required to be identical in different points across space; such difference in distributions is compensated using normalized rank transformation prior to the association analysis.

We consider two versions of the normalized rank transformation, which differ only in the direction of the ranking order, and are equivalently applicable in the context of the present study. We choose the type of the transformation so that its result may be interpreted as the empirical frequency of observing same or greater values of the indicator as compared to the current observation, calculated for a particular location on the Earth surface. The anomalies of interest (here, anomalously high values) are thus represented by small values of the transformed indicator. This choice is not essential for a single indicator, but in studies involving several indicators simultaneously (e.g. as inputs of a multi-input classifier) it may be beneficial to choose the type of the transformation for each indicator individually, depending upon the particular sign of the anomaly (high or low value), which is of interest for this indicator, so that the anomalies of interest are consistently transformed into small values for all indicators.

Physically, the stationarity assumption implies the absence of varied external conditions (or leaving the effects of such variation out of the scope of the study), and ergodicity is provided by sufficiently strong chaoticity (such as the mixing property) of the underlying dynamics. The sample size, i.e. the quantity of effectively independent (despite a possible short-range dependence) observations, must be sufficient in the sense that empirical frequencies can be seen as a valid approximation of the underlying probabilities.

If the underlying process is subject to periodic or quasiperiodic nonstationarity (generally, if it is asymptotically stationary), then part of the calculations remain meaningful, but the final result may be biased, thus, the impact of a possible nonstationarity needs to be assessed. We do that by repeating the calculation on a subset of the indicator time series obtained by omitting a part of each time period of possible nonstationarity from the full time series (here, we keep the data for September through December of each calendar year and omit the rest), and comparing the result obtained from the data with omissions to that from full data.

The applicability of other assumptions (including ergodicity of the underlying process and sufficiency of sample size) is justified by testing the final result against surrogate data simulating the null hypothesis (i.e. independence between the indicator and the event of interest), which are obtained by a time shift of the spatiotemporal data for the events of interest.

We have introduced a new formulation of local clustering coefficient for weighted correlation networks, which combines the known approaches based upon three-way partial correlations and Kendall's correlations. This modified formulation is intended to reveal more information on the underlying interaction structure as compared to the local clustering coefficient for unweighted networks, which are obtained by thresholding the pairwise correlation coefficients. Using three-way partial correlations aims at mitigating the indirect interactions effect upon the correlation network, while Kendall's correlation admits a shorter length of the sliding time window (and hence a better time resolution) when used in the context of evolving correlation networks, which capture changes in the network topology over time. This is especially important in the studies of processes occurring on time scales which are only moderately (e.g. 10 to 20 times) greater than the sampling period of the source time series used to construct a correlation network.

We have applied the above statistical methodology to compare the two mentioned local clustering coefficient formulations (calculated for an evolving correlation network of mean sea level pressure) in terms of their association to trajectories of tropical cyclone centers over the northern part of the Indian Ocean. The statistical analysis (i) confirms the association of anomalously high values of the local clustering coefficient to tropical cyclones, and (ii) shows that the newly suggested measure demonstrates a stronger association than the conventional one.

We expect the newly suggested local clustering coefficient formulation to be a powerful indicator for predicting extreme climate events, including tropical cyclones. The prediction problem is likely to benefit from a simultaneous use of several indicators, including network measures and non-network climate variables. This can be achieved e.g. by using learnable multi-input classifiers and machine learning techniques.

Our statistical methodology of association analysis can be applied to a variety of problems in climate studies, such as analyzing connections between various climate network measures (and their combinations) and particular climate events. It can also find application beyond the context of climate science, in establishing association between a spatially inhomogeneous random field and an ensemble of spatiotemporal trajectories or binary-valued spatiotemporal data.

\section*{Acknowledgement}
This research was supported by the Russian Ministry of Science and Education under Agreement No. 075-15-2020-808.

\bibliography{report}

\end{document}